\documentclass[german,english]{bvm2020}

\def\articlenumber{2773}

\date{}
\title{Prediction of MRI Hardware Failures based on Image Features using Ensemble Learning}

\titlerunning{Prediction using Ensemble Learning}

\author{Nadine Kuhnert$^1$, Lea Pfl\"uger$^1$, Andreas Maier$^1$}

\authorrunning{Kuhnert, Pfl\"uger \& Maier}

\institute{$^1$Pattern Recognition Lab, Friedrich-Alexander University Erlangen-Nuremberg}

\email{nadine.kuhnert@fau.de}

\begin{document}

%
\selectlanguage{english}
\newcommand\tabrotate[1]{\rotatebox{90}{#1\hspace{\tabcolsep}}}

\maketitle
\begin{abstract}
	In order to ensure trouble-free operation, prediction of hardware failures is essential. This applies especially to medical systems. Our goal is to determine hardware which needs to be exchanged before failing. In this work, we focus on predicting failures of 20-channel Head/Neck coils using image-related measurements. Thus, we aim to solve a classification problem with two classes, normal and broken coil.
	To solve this problem, we use data of two different levels. One level refers to one-dimensional features per individual coil channel on which we found a fully connected neural network to perform best. The other data level uses matrices which represent the overall coil condition and feeds a different neural network.
	We stack the predictions of those two networks and train a Random Forest classifier as the ensemble learner. 
	Thus, combining insights of both trained models improves the prediction results and allows us to determine the coil's condition with an F-score of 94.14\% and an accuracy of 99.09\%. 
\end{abstract}

\section{Introduction}
Reliable operation of medical diagnostic imaging systems is crucial for patient's health and healthcare provider's daily business. Thus, the goal is to prevent unplanned system downtimes by exchanging hardware before its malfunction. 
In Magnetic Resonance Imaging (MRI), one of the crucial hardware components are radiofrequency coils. Those coils can receive and transmit radiofrequency signals which determine the desired, diagnostic image. Coils contain different channels operating as separate, local receivers~\cite{maier2018medical}. In this work, we focus on coils which are specifically used for examinations of the human body's head and neck area.
Image features and related measurements depend next to the depicted tissue also on the coil's condition. Before every clinical scanning process, noise and signal measurements are performed from which we derive 1-D features for every coil channel as well as matrix features per coil.

The task of classification is well known in literature, described as assigning an input vector to one out of a set of discrete, predefined classes~\cite{Witten2016}, and applied to various applications~\cite{maier2019gentle}. E.g. Kuhnert et al.~\cite{kuhnert2017classification} found Neural Networks to achieve highest accuracy in classifying the examined body parts using MRI acquisition parameters. Next to Neural Networks, also Random Forest (RF) Classifier is widely used and has been introduced by~\cite{breiman2001random}. Random Forests combine tree predictors where the search space is a randomly chosen subset. 
Often, several machine learning techniques train an ensemble of models to leverage their results in combination~\cite{Witten2016}. One form of ensemble learning is stacking which was introduced by Wolpert~\cite{Wolpert92} and refers to a meta learner which is trained on the reliability of base classifiers as inputs.

Several ML techniques have been applied to address the task of hardware failure prediction as described in~\cite{Chigurupati2016}. For MRI in specific, Jain et al.~\cite{jain2019image} showed that defective components can be identified using image features. 

\section{Materials and methods}
Our approach classifies image features and consecutively Head/Neck coils of MRI systems as normal or defective. We trained different models on measurements per coil channel and features per coil, individually. In the following, we describe the underlying data, preprocessing, training of our classifiers and the final ensemble learning step in order receive one consolidated prediction per coil.
\subsection{Data}
All data sets were acquired by 238 Siemens MAGNETOM Aera 1.5T MRI systems using a 20-channel Head/Neck coil. We enrich those data with measurements of normal and broken coils performed at Siemens' research halls. 
During coil adjustment measurements we collect parameters which represent image features.
Thus, we work only on entirely anonymized, non-clinical data.

The collected parameters contain two degrees. On the one hand side, we extract four numerical, one-dimensional features per coil channel. Those features are calculated for all 329338 samples and depict the noise level, signal to noise ratio (CSP), signal ratio of body coil to Head/Neck coil and the ratio of CSP to channel signal in the isocenter. On the other hand side, we use the Noise Covariance Matrix (NCM) as a feature for the entire coil. It holds the covariances of all individual noise values per channel resulting in 19687 matrices.

We perform leave-several-coils-out cross validation to calculate all performance measures on disjoint coil sets for fitting and testing. Furthermore, we apply a stratified split on the fitting data set and use 70\% for base model training and 30\% for hyper-parameter tuning. 
On average, one fold contains 29985 one-dimensional features and 1791 matrices.

\subsection{Preprocessing}
We performed normalization and augmentation. All features were normalized by subtracting the means and scaling to unit variance. 
To overcome the imbalanced class distribution (6.8\% broken coils matrices), we augmented non-testing NCM instances. Thus, we permuted on-diagonal elements randomly. The off-diagonal elements were placed according to their respective on-diagonal element and thus, the matrices' rows and columns keep their relations. This results in N-1 generated matrices for every NxN matrix entered, where N equals 20 in our case.

\subsection{Base learner training}
Prediction of coils as normal or defective was achieved using different classification methods on one-dimensional features and matrices, individually.

We applied a Fully Connected Neural Network (FCN) to the numerical features. It consists of four blocks, where each block is composed of one fully connected (FC) layer. This is followed by a batch normalization and a dropout operation to prevent over-fitting. A rectified linear unit (ReLU) activation function is used for all FC layers, only the number of elements is changed within the
different FC layers. Parameters and dropout rate were found using hyper-parameter tuning. The result of the four blocks is finally fed to a last FC layer with two elements. As an activation function we used softmax.

Secondly, we trained a Convolutional Neural Network (CNN) on the available NCM input. We tried four different CNN configurations. Please see the exact parameters and structures of CNN1 and CNN2 in Table~\ref{tab:NN_config}. CNN4 is similarly built to CNN2 but has no padding included and contains only one convolutional layer before and after the first MaxPooling layer. CNN3 is comparable to CNN4 but has three dropout layers included.
\begin{table}[t]
	\caption{Configuration details of CNN1 and CNN2 used for training on matrix features.}
	\centering
	\resizebox{\textwidth}{!}{%
		
		\begin{tabular*}{\textwidth}{l@{\extracolsep\fill}ccccccccc}
			\hline
			{Layer} &  {1} &  {2} &  {3} &  {4} &  {5} &  {6} &  {7} &  {8} &  {9} \\ 
			\hline
			
			\tabrotate{{CNN1}} & \begin{tabular}[c]{@{}c@{}}Conv2D, \\ 6, \\ 'relu', \\ (3,3), \\ padding \\= 'same'\end{tabular} & \begin{tabular}[c]{@{}c@{}}Average\\Pooling \end{tabular}& \begin{tabular}[c]{@{}c@{}}Conv2D, \\ 16, \\ 'relu', \\ (3,3), \\ padding \\= 'same'\end{tabular} & \begin{tabular}[c]{@{}c@{}}Average\\Pooling \end{tabular} & Flatten & \begin{tabular}[c]{@{}c@{}}Dense, \\ 64, \\ 'relu'\end{tabular} & \begin{tabular}[c]{@{}c@{}}Dense, \\ 2, \\ 'soft-\\max'\end{tabular} &  &    \\ \hline
			
			\tabrotate{{CNN2}} & \begin{tabular}[c]{@{}c@{}}Conv2D, \\ 16, \\ 'relu', \\ (3,3), \\ padding \\= 'same'\end{tabular} & \begin{tabular}[c]{@{}c@{}}Conv2D, \\ 16, \\ 'relu', \\ (3,3), \\ padding \\= 'same'\end{tabular} & \begin{tabular}[c]{@{}c@{}}Max-\\Pooling,\\ (2,2)\end{tabular} & \begin{tabular}[c]{@{}c@{}}Conv2D, \\ 32, \\ 'relu', \\ (3,3), \\ padding \\= 'same'\end{tabular} & \begin{tabular}[c]{@{}c@{}}Conv2D, \\ 32, \\ 'relu', \\ (3,3), \\ padding \\= 'same'\end{tabular} & \begin{tabular}[c]{@{}c@{}}Max-\\Pooling,\\ (2,2)\end{tabular} & Flatten & \begin{tabular}[c]{@{}c@{}}Dense, \\ 64, \\ 'relu'\end{tabular} & \begin{tabular}[c]{@{}c@{}}Dense, \\ 2, \\ 'soft-\\max'
			\end{tabular} \\
			
			\hline
			
		\end{tabular*}%
	}
	\label{tab:NN_config}
\end{table}
Table~\ref{tab:resMatrix_classification} holds the average prediction performance measures, accordingly. Model CNN1 reached the highest performance with a F-score of 66.00\% for all tested cross-validation splits,  model CNN3 lowest with 29.29\%. CNN4 almost reaches as good results as CNN1.  
\begin{table}[t]
	\caption{Average prediction performance measures after 10-fold cross-validation  for the different CNNs applied to NCMs.}
	\centering
	\begin{tabular*}{\textwidth}{l@{\extracolsep\fill}cccc}
		\hline
		\% & {Accuracy} & {Precision} & {Recall} & {F-Score}  \\ 
		{CNN1} & 94.82 & 97.99 & 58.06 & 66.00 \\
		{CNN2} & 92.47 & 75.92& 31.39& 36.25 \\
		{CNN3} & 92.45 & 100.00& 20.05 & 29.29 \\
		{CNN4} & 93.41 & 94.62 & 51.56 & 60.28 \\
		\hline
	\end{tabular*} 
	\label{tab:resMatrix_classification}
\end{table}

\subsection{Ensemble learner training}
Finally, in order to leverage all information and decide for the overall predicted class, we stacked the predictions of two base learners, FCN and CNN2, and applied a RF classifier. We extracted four different features from the two base learners and fed them to the meta learner. 
We aggregate the FCN prediction results on coil element level, by calculating the minimum, standard deviation and mean prediction values on coil level. Furthermore, we incorporate the CNN2 prediction probability per coil as forth feature. 
50\% of those extracted feature instances are used for fitting the RF model. We tested the model on the remaining 50\% instances. This enables one consolidated prediction result per coil.

\section{Results}
Main achieved results are presented for the individual models next to the combined model. 

\subsection{Base learner results 1D}
We achieved an average prediction F-score of 82.43\% after 10-fold cross-validation by applying the described FCN to our original set of one-dimensional features on coil element level. Table~\ref{tab:res1d_NN} shows the respective achieved accuracy of 99.74\% next to precision and recall both holding values above 82\%. For more details, Table~\ref{tab:res1d_NN} also presents the confusion matrix of the FCN showing True Negative (TN), False Positive (FP), False Negative (FN) and True Positive (TR) rates and positive (P) and negative (N) samples in percent.
\begin{table}[t]
	\caption{Average prediction performance measures and confusion matrix of FCN on coil element level after 10-fold cross-validation.}
	\centering
	\begin{tabular*}{\textwidth}{l@{\extracolsep\fill}cccccccccc}
		\hline
		{\%} & {Accuracy} & {Precision} & {Recall} & {F-Score} & {TN} & {FP} & {FN} & {TP} & {N} & {P}\\ 
		{FCN} 	& 99.74 & 86.12 & 82.37 & 82.43 & 99.93 & 0.07 & 17.61 & 82.38 & 97.6 & 2.4\\
		\hline
	\end{tabular*} 
	\label{tab:res1d_NN}
\end{table}

\subsection{Base learner results matrix}
By augmenting the matrix data set, we achieved the desired ratio of 20\% for number of defective vs. normal samples. We gained significant improvement in prediction performance after augmentation. Thus, all further results shown in this section are based on  training with artificial instances added during preprocessing. Table \ref{tab:NCM_sampling_acc} presents the performances of the four CNNs applied to our data set including artificial instances next to the respective confusion matrix.
\begin{table}[ht]
	\centering
	\begin{tabular}{lcccc|cccc}
		\hline
		{\%} & {Accuracy} & {Precision} & {Recall} & {F-Score} & {TN} & {FP} & {FN} & {TP} \\ 
		{CNN1} & 98.12 & 99.13 & 85.74 & 91.25 & 99.85& 0.15 & 14.25 & 85.75 \\
		{CNN2} & {98.27} & 99.45& {86.49} & {91.86} & 99.91 & 	0.09 & 13.50 & 86.50\\
		{CNN3} & 97.61 & {99.90}& 82.33 & 88.83 & 99.91 &	0.09& 17.66 & 82.34\\
		{CNN4} & 98.09 & 99.52 & 85.30 & 91.05 & 99.92 & 0.08 & 14.69 & 85.31 \\
		\hline
	\end{tabular}
	\caption{Average prediction performance measures for CNNs applied to data set with augmentation.}
	\label{tab:NCM_sampling_acc}
\end{table}

\subsection{Ensemble learner results}
\begin{figure}[ht]
	\centering
	\includegraphics[width=0.9\textwidth]{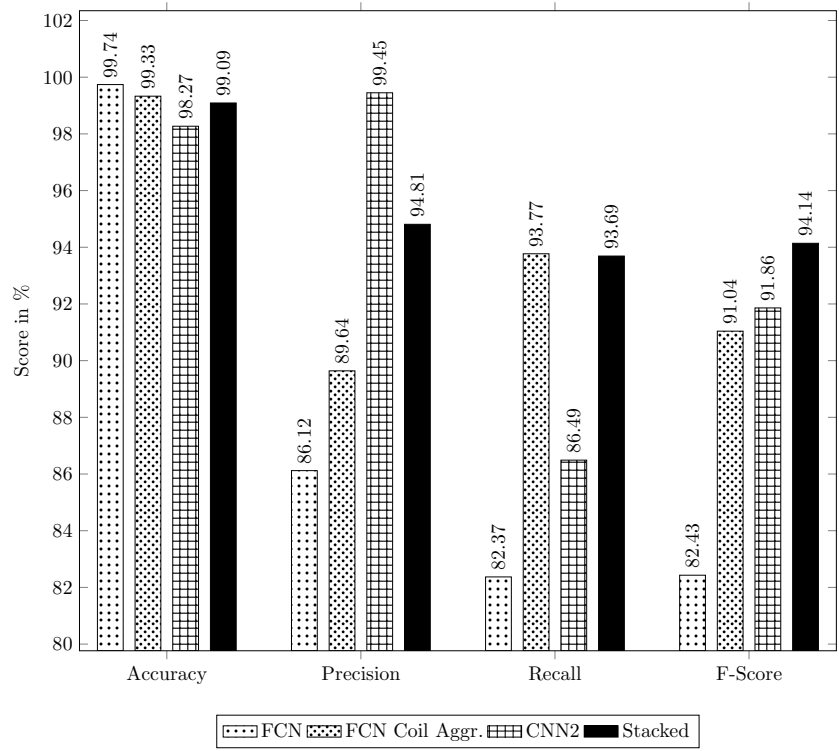}
	\caption{Average prediction scores for the two base classifiers FCN and CNN2 next to FCN aggregated per coil and the stacked meta learner.}
	\label{fig:results_stacking}
\end{figure}

Through aggregating the FCN results and only using their minimal prediction probability per coil, we observe an increase in performance from an average F-score of 82.43\% to 91.04\%. The average F-score for the best NCM classification lies at 91.86\%. By combining the predictions of those two
base learners and training a Decision
Tree on these predicted instances we reach a final F-score of 94.14\% and accuracy of 99.09\%. The individual model performances are visualized in Fig.~\ref{fig:results_stacking}.

\section{Discussion}
We achieve satisfying results by training a FCN on features per coil element and aggregating the individual decision to a prediction per entire coil. The aggregation is necessary to overcome correlations of broken and normal elements within one coil affecting normal channels and their measurements such that they lead to false positives. 
In case of training a model on our matrix features, augmentation of data allowed to improve the classification result drastically due to avoiding overfitting to specific coil elements. As the prediction should not be affected by the location of the highest on-diagonal element value, augmentation was necessary to overcome unrealistic correlation of broken coil to specific coil elements. 
By combining the learnings of the two trained models in a final ensemble learning step, we could leverage the two different view points and information degrees of our data. Thus, stacking improved the resulting performance measures even more. 

To our knowledge, we are the first ones to combine coil features with coil channel specific information using ensemble learning for the application of hardware failure prediction. For further research, the method should be expanded to different coil types and even larger data sets. As this approach is planned to be integrated into the Quality Assurance procedure, a next step is to also incorporate a feedback loop from technicians validating the model's prediction for self-learning improvement.

\bibliographystyle{bvm2020}

\bibliography{2773}
\marginpar{\color{white}E\articlenumber} 
\end{document}